\documentclass[a4paper]{revtex4}
\usepackage{graphicx}
\usepackage{fancyhdr}
\usepackage{amsmath}
\usepackage{color}

\begin{document}

\title{Enhanced $hVV$ couplings in the Georgi-Machacek model and beyond}


\author{Heather E.~Logan}
%\email{logan@physics.carleton.ca}
\affiliation{Ottawa-Carleton Institute for Physics, Carleton University, 
%1125 Colonel By Drive, 
Ottawa, Ontario K1S 5B6 Canada}

\begin{abstract}
In this talk I discuss extended Higgs sectors in which the 125~GeV Higgs boson couplings to $W$ and $Z$ bosons can be larger than in the Standard Model.  Constraints from perturbative unitarity and the electroweak rho parameter limit the number of possible models to a tractable few.  Focusing on generalizations of the Georgi-Machacek model and taking advantage of the custodial symmetry, I show that existing experimental and theoretical constraints can be combined to set an upper limit on the enhancement of the Higgs couplings to $W$ and $Z$ bosons.  This talk is based mostly on Ref.~\cite{Logan:2015xpa}.
%arXiv:1502.01275.
\end{abstract}

\maketitle

\thispagestyle{fancy}


%%%%%%%%%%%%%%%%%%%%%%%%%%%%%%%%%%
\section{Motivation: boosting the $hVV$ coupling \label{sec:prelims}}

With the discovery of a Standard Model (SM)-like Higgs boson at the CERN Large Hadron Collider (LHC), measurement of the properties of the Higgs have become a top priority.  The LHC measures rates in particular Higgs production and decay channels,
\begin{equation}
	{\rm Rate}_{ij} = \frac{\sigma_i \Gamma_j}{\Gamma_{\rm tot}}
	= \frac{\kappa_i^2 \sigma_i^{\rm SM} \cdot \kappa_j^2 \Gamma_j^{\rm SM}}
	{\sum_k \kappa_k^2 \Gamma_k^{\rm SM} + \Gamma_{\rm new}},
\end{equation}
where $\kappa_i$ is the ratio of the Higgs coupling to SM particles $i$ normalized to its SM-predicted value and $\Gamma_{\rm new}$ is the Higgs decay width into new, non-SM final states.

If all the $\kappa_i$ are equal with $\kappa_i^2 \equiv \kappa^2 = 1/(1 - {\rm BR}_{\rm new}) \geq 1$, all Higgs signal rates at the LHC will be identical to their SM predicted values~\cite{Zeppenfeld:2000td}.  In this case a coupling enhancement will hide the presence of new decays, and new decays will hide the presence of a coupling enhancement! (Measurements of off-shell $gg \ (\to h^*) \to ZZ$~\cite{Kauer:2012hd} assuming no new resonances in the $s$-channel can constrain $\kappa$; however, a light second resonance $H$ can cancel the effect of modified $h$ couplings in this process~\cite{Logan:2014ppa}.)  This LHC ``blind spot'' motivates us to study explicit models with $\kappa > 1$ to gain insight into how this scenario can be constrained.

%%%%%%%%%%%%%%%%%%%%%%%%%%%%%%%%%%

The first interesting feature when considering models in which the $h$ coupling to $W$ or $Z$ boson pairs is larger than in the SM is the unitarization of longitudinally-polarized vector boson scattering at high energies $E \gg v$, where $v \simeq 246$~GeV is the SM Higgs vacuum expectation value (vev).  In the SM, Higgs exchange famously cancels the $E^2/v^2$ growth of the $W^+_LW^-_L \to W^+_LW^-_L$ amplitude~\cite{Lee:1977yc}.  In doublet or singlet extensions of the SM, this cancellation implies a sum rule $(\kappa_V^h)^2 + (\kappa_V^H)^2 = 1$~\cite{Gunion:1990kf}.

When $\kappa_V^h > 1$, it is not possible to satisfy such a sum rule with only neutral Higgs exchanges; instead, one needs a doubly-charged Higgs boson exchanged in the $u$-channel~\cite{Falkowski:2012vh,Gunion:1989we}.  The presence of a doubly-charged Higgs coupled to SM gauge bosons implies the presence of scalars in isospin representations larger than the usual doublet.  This can also be seen straightforwardly in an explicit model implementation of $hWW$ couplings larger than in the SM.  In particular, a coupling enhancement requires a scalar with isospin one or larger, which must have a non-negligible vev and must mix into the observed Higgs boson $h$.

How large can the isospin of such a scalar be?  The maximum isospin is constrained by the requirement that the weak charges not become too large, thereby violating tree-level perturbative unitarity of $V_T V_T \to \phi\phi$ scattering amplitudes, where $V_T$ are \emph{transversely} polarized SU(2)$_L$ gauge bosons and $\phi$ are the new scalars.  The general result for the zeroth partial wave amplitude for this process for a complex scalar multiplet with isospin $T = (n-1)/2$ is~\cite{Hally:2012pu}
\begin{equation}
	a_0^{\rm max} = \frac{g^2}{16 \pi} \frac{(n^2-1) \sqrt{n}}{2 \sqrt{3}}.
\end{equation}
Requiring that this amplitude satisfy the usual perturbativity constraint $|{\rm Re} \, a_0| < 1/2$,  one obtains an upper bound on the isospin $T \leq 7/2$ for a complex scalar multiplet. (For a real multiplet this becomes $T \leq 4$ due to the smaller multiplicity of scalars.)  The constraints become even tighter if more than one large scalar multiplet is present because of the higher multiplicity of final states.

Of course, adding scalars in SU(2)$_L$ representations larger than the doublet leads to well-known problems with the electroweak $\rho$ parameter, which is the ratio of the strengths of the neutral and charged weak currents.  For a general Higgs extension, $\rho$ can be written as~\cite{Gunion:1989we}
\begin{equation}
	\rho = \frac{M_W^2}{M_Z^2 \cos\theta_W}
	= \frac{\sum_k 2[T_k (T_k + 1) - Y_k^2/4] v_k^2}
		{\sum_k Y_k^2 v_k^2},
	\label{eq:rho}
\end{equation}
where the scalar vevs are defined as $\langle \phi_k^0 \rangle = v_k/\sqrt{2}$ ($\langle \phi_k^0 \rangle = v_k$) for a complex (real) representation and the hypercharge operator is normalized according to $Q = T^3 + Y/2$.  Given the extremely tight experimental constraint $\rho = 1.00040 \pm 0.00024$~\cite{Agashe:2014kda}, a convincing extension of the SM containing scalars with isospin one or larger and non-negligible vevs should address this constraint via some clever model-building.

There are two approaches to adding vev-carrying scalars with isospin one or larger while preserving the $\rho$ parameter.  The first takes advantage of the fact that certain combinations of isospin and hypercharge in Eq.~(\ref{eq:rho}) happen to yield $\rho = 1$ automatically.  After the usual doublet, the next-simplest solution is the septet with $T = 3$, $Y = 4$~\cite{Hisano:2013sn,Kanemura:2013mc}.  (The third solution, with $T = 25/2$, $Y = 15$, is forbidden by tree-level perturbative unitarity of the weak charges, as are all higher solutions~\cite{Hally:2012pu}.)
The disadvantage of the septet extension of the SM is that the renormalizable scalar potential contains an accidental global U(1) symmetry under which the septet is free to rotate, so that one cannot give the septet a vev through spontaneous symmetry breaking without generating a physical massless Goldstone boson.  This was dealt with in Ref.~\cite{Hisano:2013sn} by coupling the septet to the SM Higgs doublet through a dimension-7 operator of the form $X \Phi^* \Phi^5$, where $X$ is the septet and $\Phi$ is the doublet.  This is fine, and Ref.~\cite{Hisano:2013sn} gives an explicit example of an ultraviolet (UV) completion, but it implies that the UV completion must be nearby.

The second approach is to impose a global SU(2)$_L \times $SU(2)$_R$ symmetry on the scalar sector~\cite{Georgi:1985nv}.  This global symmetry breaks to custodial SU(2) upon electroweak symmetry breaking, thereby preserving $\rho = 1$ at tree level.  The disadvantage of this class of extensions of the SM, the simplest of which is known as the Georgi-Machacek model, is that the global SU(2)$_R$ symmetry is explicitly broken by the gauging of hypercharge~\cite{Gunion:1990dt}.  The special relations among the parameters of the full gauge-invariant scalar potential that are imposed by the global SU(2)$_R$ symmetry can thus only hold at one energy scale, because they are violated by the running due to hypercharge~\cite{Garcia-Pepin:2014yfa}.  Again, this implies that the UV completion that is responsible for the approximate SU(2)$_R$ global symmetry in the scalar sector must be nearby.

Even given the theoretical caveats, these scalar sector extensions can lead to very interesting phenomenology, and are worth studying for that reason alone.  In this talk I focus on the Georgi-Machacek model and its generalizations to higher isospin.

%%%%%%%%%%%%%%%%%%%%%%%%%%%%%%%%%%
\section{The Georgi-Machacek model and its generalizations \label{sec:models}}

The Georgi-Machacek model~\cite{Georgi:1985nv} adds one real isospin triplet $\xi$ and one complex isospin triplet $\chi$ to the SM Higgs sector.  The scalar field content can be written in the form
\begin{equation}
	\Phi = \left( \begin{array}{cc}
	\phi^{0*} &\phi^+  \\
	-\phi^{+*} & \phi^0  \end{array} \right), \qquad \qquad
	X =	\left(
	\begin{array}{ccc}
	\chi^{0*} & \xi^+ & \chi^{++} \\
	 -\chi^{+*} & \xi^{0} & \chi^+ \\
	 \chi^{++*} & -\xi^{+*} & \chi^0  
	\end{array}
	\right),
\end{equation}
which displays explicitly the transformation properties under the SU(2)$_L \times$SU(2)$_R$ global symmetry (for a review and complete list of references current as of April 2014, see Ref.~\cite{Hartling:2014zca}).  

The physical spectrum is dictated almost entirely by the custodial symmetry.  The bidoublet $\Phi$ and bitriplet $X$ decompose under SU(2)$_L \times$SU(2)$_R \to $~SU(2)$_{\rm custodial}$ according to
\begin{equation}
	{\rm Bidoublet:} \ 2 \otimes 2 \to 3 \oplus 1, \qquad \qquad 
	{\rm Bitriplet:} \ 3 \otimes 3 \to 5 \oplus 3 \oplus 1.
\end{equation}
The two custodial singlets mix, yielding CP-even neutral mass eigenstates $h$ and $H$.  The two custodial triplets mix by an angle controlled by the vevs of the doublet and the triplets, yielding a physical custodial triplet $(H_3^+, H_3^0, H_3^-)$ (with all members degenerate in mass) and the usual triplet of Goldstone bosons.  The custodial fiveplet $(H_5^{++}, H_5^+, H_5^0, H_5^-, H_5^{--})$ (again with all members degenerate in mass) contains the new states responsible for unitarizing longitudinal vector boson scattering in the presence of $hVV$ couplings larger than in the SM.  The phenomenology of the Georgi-Machacek model has been increasingly investigated in recent years.

In this talk my focus is on the generalizations of the Georgi-Machacek models to higher isospin~\cite{Galison:1983qg,Logan:2015xpa}.  By replacing the bitriplet $X$ with a ``bi-$n$-plet'' one can construct a series of models (denoted ``GGM$n$'') with the following scalar field content under SU(2)$_L \times$SU(2)$_R \to $~SU(2)$_{\rm custodial}$:
\begin{equation}
	{\rm Bidoublet:} \ 2 \otimes 2 \to 3 \oplus 1, \qquad \qquad
	\left\{ \begin{array}{r l}
	{\rm Bitriplet:} \ 3 \otimes 3 \to 5 \oplus 3 \oplus 1 &\quad {\rm (GM)}, \\
	{\rm Biquartet:} \ 4 \otimes 4 \to 7 \oplus 5 \oplus 3 \oplus 1 &\quad {\rm (GGM4)}, \\
	{\rm Bipentet:} \ 5 \otimes 5 \to 9 \oplus 7 \oplus 5 \oplus 3 \oplus 1 &\quad {\rm (GGM5)}, \\
	{\rm Bisextet:} \ 6 \otimes 6 \to 11 \oplus 9 \oplus 7 \oplus 5 \oplus 3 \oplus 1 &\quad {\rm (GGM6)}.
	\end{array} \right.
\end{equation}
Larger bi-$n$-plets are forbidden by the perturbativity of the weak charges.  Again, the physical spectrum is dictated almost entirely by the custodial symmetry.  It again contains two (mixed) custodial singlets $h$ and $H$, a physical custodial triplet $(H_3^+, H_3^0, H_3^-)$, and a custodial fiveplet $(H_5^{++}, H_5^+, H_5^0, H_5^-, H_5^{--})$ which contains the new states responsible for unitarizing longitudinal vector boson scattering.  The generalized models also contain additional scalars in larger custodial multiplets, whose phenomenology I will henceforth ignore.

%%%%%%%%%%%%%%%%%%%%%%%%%%%%%%%%%%
\section{Phenomenology \label{sec:pheno}}

The phenomenology of the custodial singlets, triplet, and fiveplet can be constructed for the generalized Georgi-Machacek models completely in parallel to the construction in the original Georgi-Machacek model.

The vevs of the bidoublet and bi-$n$-plet are defined as 
$\langle \Phi \rangle = (v_{\phi}/\sqrt{2}) I_{2\times 2}$ and $\langle X \rangle = v_n I_{n\times n}$,
which we can parameterize in terms of a mixing angle
$c_H \equiv \cos\theta_H = v_{\phi}/v$.  

%\subsection{Custodial singlets}

The two custodial-singlet states are mixtures of $\phi^{0,r} \equiv \sqrt{2} \, {\rm Re} \, \phi^0$ and the custodial singlet from $X$, which we call $H_1^{\prime 0}$ (explicit expressions are given in Ref.~\cite{Logan:2015xpa}):
\begin{equation}
	h = c_{\alpha} \phi^{0,r} - s_{\alpha} H_1^{\prime 0}, \qquad \qquad
	H = s_{\alpha} \phi^{0,r} + c_{\alpha} H_1^{\prime 0},
\end{equation}
where $c_{\alpha}$ and $s_{\alpha}$ are the cosine and sine of a second mixing angle $\alpha$.  The couplings of $h$ and $H$ to vector boson pairs and to fermion pairs are given relative to the corresponding couplings of the SM Higgs boson by
\begin{eqnarray}
	\kappa_V^h &=& c_{\alpha} c_H - \sqrt{A} \, s_{\alpha} s_H, \qquad 
		\kappa_f^h = c_{\alpha}/c_H, \nonumber \\
	\kappa_V^H &=& s_{\alpha} c_H + \sqrt{A} \, c_{\alpha} s_H, \qquad
		\kappa_f^H = s_{\alpha}/c_H,
\end{eqnarray}
where the $\sqrt{A}$ factor comes from the SU(2)$_L$ generators acting on the larger multiplets and is given by $A = 4 T (T+1)/3$.  For the Georgi-Machacek model and its three generalizations we have
\begin{equation}
	A_{\rm GM} = 8/3, \qquad 
	A_{\rm GGM4} = 15/3, \qquad
	A_{\rm GGM5} = 24/3, \qquad
	A_{\rm GGM6} = 35/3.
\end{equation}
Note also that $\kappa_V^h \leq [1 + (A-1) s_H^2]^{1/2}$, and this upper bound is saturated for fixed $s_H$ when $\kappa_V^H = 0$.  Because $A > 1$ in these models, large enhancements of $\kappa_V^h$ are possible for large $s_H$, up to a maximum of about 3.3 in the GGM6 model (perturbativity of the top quark Yukawa coupling forces $\tan\theta_H < 10/3$~\cite{Barger:1989fj}).

%\subsection{Custodial triplet}

The phenomenology of the custodial triplet in the generalized Georgi-Machacek models can be expressed in a way that makes it identical to that of the original Georgi-Machacek model.  The two custodial triplets are mixtures of $\Phi_3 \equiv (\phi^+, \phi^{0,i} \equiv \sqrt{2} \, {\rm Im} \, \phi^0, \phi^{+*})$ and the custodial triplet $H_3^{\prime}$ from $X$:
\begin{equation}
	G^{0, \pm} = c_H \Phi_3^{0,\pm} + s_H H_3^{\prime 0, \pm}, \qquad \qquad
	H_3^{0, \pm} = -s_H \Phi_3^{0,\pm} + c_H H_3^{\prime 0, \pm}, 
\end{equation}
where $G^{0, \pm}$ are the Goldstone bosons eaten by the $Z$ and $W^{\pm}$ and $H_3^{0, \pm}$ are the physical custodial-triplet scalars.  These states are vector-phobic, with no tree-level couplings of the form $H_3 VV$.

The couplings of the $H_3^{0,\pm}$ to fermions are completely analogous to those of the pseudoscalar $A^0$ and charged Higgs pair $H^{\pm}$ of the Type-I two-Higgs-doublet model, with Feynman rules given by
\begin{equation}
	H_3^0 \bar u u: \ \frac{m_u}{v} \tan \theta_H \gamma_5, \qquad \quad
	H_3^0 \bar d d: \ -\frac{m_d}{v} \tan \theta_H \gamma_5, \qquad \quad
	H_3^+ \bar u d: \ -i \frac{\sqrt{2}}{v} V_{ud} \tan\theta_H
		\left( m_u P_L - m_d P_R \right).
\end{equation}
The $Z H_3^+ H_3^-$ coupling is also the same as the $Z H^+ H^-$ coupling in the Type-I two-Higgs-doublet model, so that the constraints from $b \to s \gamma$, $B_s \to \mu \mu$, $R_b$, etc., can be taken over directly from the Georgi-Machacek model~\cite{Hartling:2014aga}.  

Indeed, it should be possible to define a mapping that enables the LHC constraints on the Type-I two-Higgs-doublet model to be taken over to the custodial triplet of the Georgi-Machacek model and its generalizations.  We leave this to future work.

%\subsection{Custodial fiveplet}

The phenomenology of the custodial fiveplet can also be constructed in parallel to that of the original Georgi-Machacek model.  Because the custodial fiveplet is made up entirely of the states of the bi-$n$-plet, it does not couple to fermions.  Instead, the $H_5VV$ couplings are nonzero, and are very different from any two-Higgs-doublet phenomenology.  The custodial symmetry fixes the $H_5VV$ Feynman rules to have a common form in all the models:
\begin{eqnarray}
	H_5^0 W^+_{\mu} W^-_{\nu}: &\ & - i \frac{2 M_W^2}{v} \frac{g_5}{\sqrt{6}} g_{\mu\nu},
		\qquad \qquad \qquad 
	H_5^0 Z_{\mu} Z_{\nu}: \  i \frac{2 M_Z^2}{v} \sqrt{\frac{2}{3}} g_5 g_{\mu\nu},
		\nonumber \\
	H_5^+ W_{\mu}^- Z_{\nu}: &\ & - i \frac{2 M_W M_Z}{v} \frac{g_5}{\sqrt{2}} g_{\mu\nu},
		\qquad \qquad
	H_5^{++} W_{\mu}^- W_{\nu}^-: \  i \frac{2 M_W^2}{v} g_5 g_{\mu\nu},
\end{eqnarray}
where the common factor $g_5$ is fixed by the longitudinal vector boson unitarization sum rule~\cite{Falkowski:2012vh,Gunion:1989we},
\begin{equation}
	(\kappa_V^h)^2 + (\kappa_V^H)^2 - \frac{5}{6} (g_5)^2 = 1, \qquad {\rm or} \qquad
	(\kappa_V^h)^2 \leq 1 + \frac{5}{6} (g_5)^2.
	\label{eq:kvhrule}
\end{equation}
This sum rule is a result of the custodial symmetry in the scalar sector, and holds for all the generalized Georgi-Machacek models.

We will take advantage of this sum rule together with the direct-search constraints on the custodial fiveplet states to constrain the maximum enhancement of $\kappa_V^h$ in this class of models.

%%%%%%%%%%%%%%%%%%%%%%%%%%%%%%%%%%%%%%%%%%%%
\section{Constraints}

We focus on constraining the $H_5VV$ coupling strengths as a function of the common $H_5$ mass.  This will allow us to set upper bounds on the $hVV$ coupling as a function of the $H_5$ mass via the second expression in Eq.~(\ref{eq:kvhrule}).

\begin{figure}
\resizebox{0.5\textwidth}{!}{\includegraphics{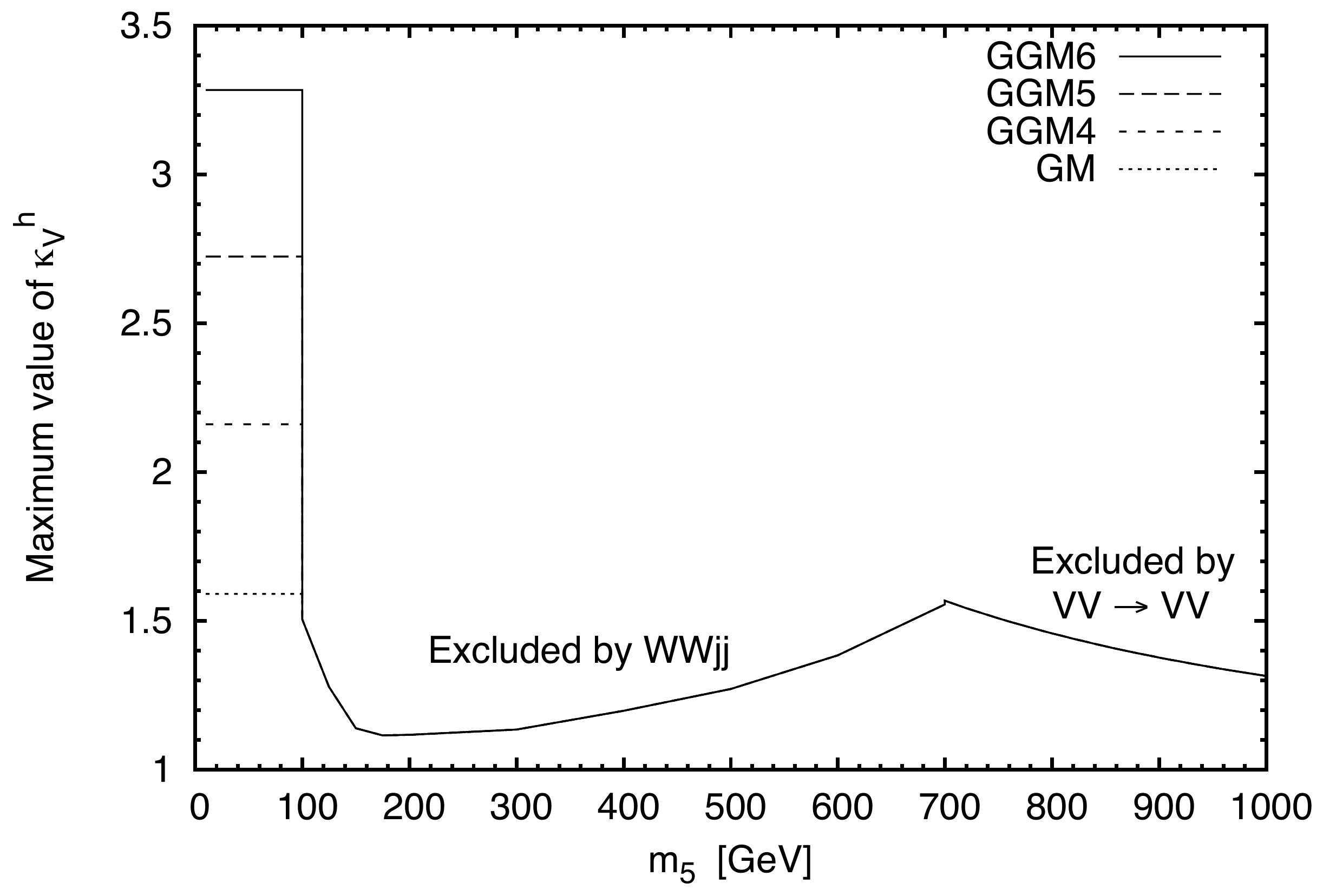}}%
\resizebox{0.5\textwidth}{!}{\includegraphics{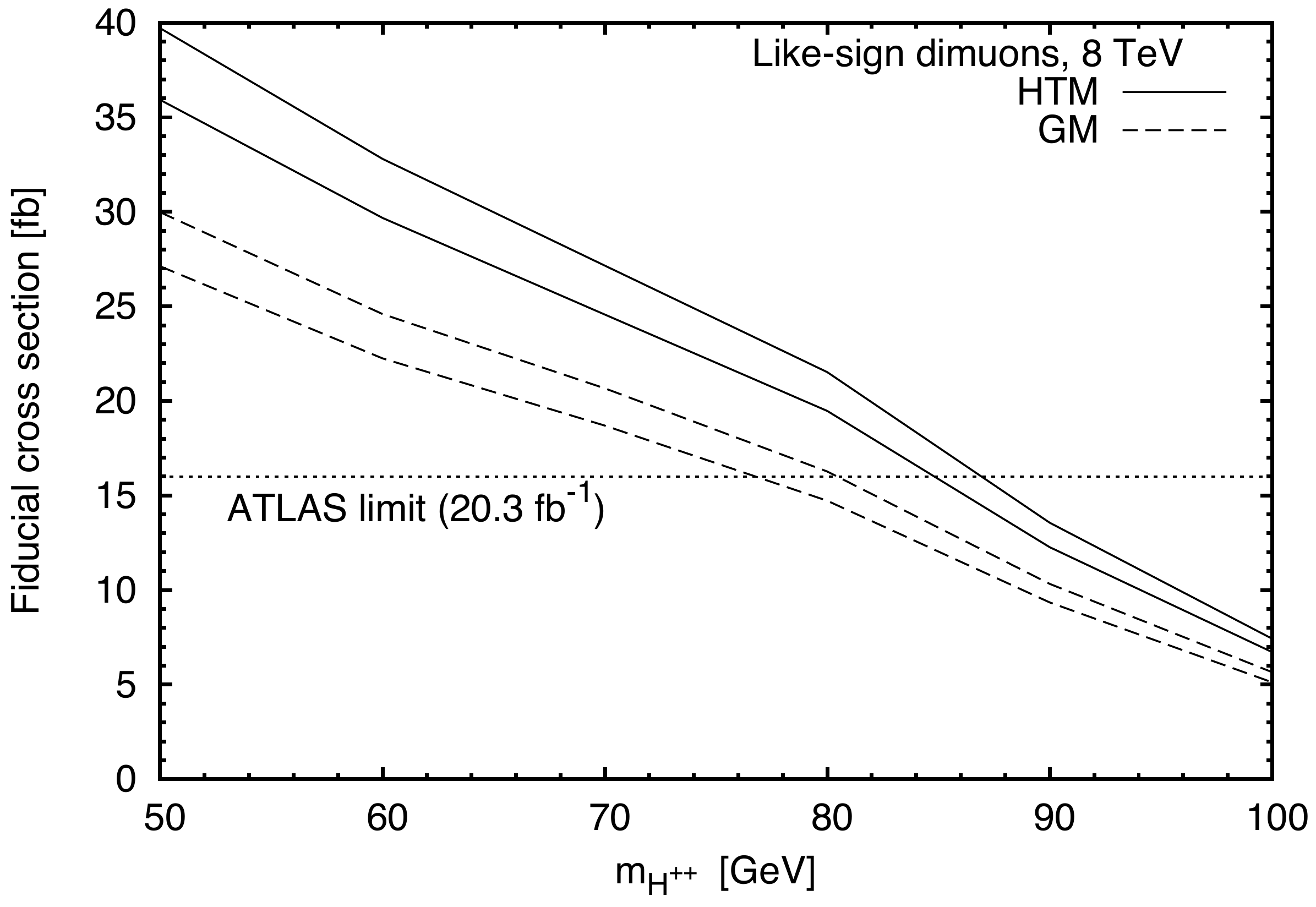}}
\caption{Left: Upper bound on $\kappa_V^h$ as obtained from the sum rule in Eq.~(\ref{eq:kvhrule}) as a function of the common $H_5$ mass $m_5$.  Constraints shown are from the like-sign $WWjj$ cross section~\cite{Chiang:2014bia} and perturbative unitarity of longitudinal $VV$ scattering amplitudes~\cite{Logan:2015xpa}.  
Right: Cross sections after cuts for like-sign dimuon production from $H^{\pm\pm}$ decay in the Higgs Triplet Model (solid lines)~\cite{Kanemura:2014ipa} and the generalized Georgi-Machacek models (dashed lines)~\cite{Logan:2015xpa}, together with the ATLAS cross section limit (horizontal dotted line)~\cite{ATLAS:2014kca}.
Both plots are from Ref.~\cite{Logan:2015xpa}.  \label{fig1}}
\end{figure}

Our first constraint is from the ATLAS measurement of the like-sign $WW$ cross section in association with two jets in a vector boson fusion topology~\cite{ATLASWWjj}.  This measurement has been recast in Ref.~\cite{Chiang:2014bia} as an upper bound on the additional production cross section for like-sign $W$ boson fusion to $H_5^{\pm\pm}$ followed by decays back to like-sign $W^{\pm}W^{\pm}$.  This excludes large $H_5VV$ coupling strengths, and hence large $hVV$ coupling enhancements, for $H_5$ masses in the range 100--700~GeV, as shown in the left panel of Fig.~\ref{fig1}.

We can obtain a second constraint for high $H_5$ masses from the requirement of tree-level perturbative unitarity of the \emph{energy-independent} part of the longitudinal $VV \to VV$ scattering amplitudes.  This requirement leads to the famous upper bound on the Higgs boson mass in the SM~\cite{Lee:1977yc}, $m_{h_{\rm SM}}^2 < 16 \pi v^2 / 5 \simeq (780~{\rm GeV})^2$, where we have included contributions from $WW \to WW$, $WW\to ZZ$, and $ZZ \to ZZ$.  In the generalized Georgi-Machacek models this bound becomes
\begin{equation}
	\left[ (\kappa_V^h)^2 m_h^2 + (\kappa_V^H)^2 m_H^2 + \frac{2}{3} (g_5)^2 m_5^2 \right] 
	< \frac{16 \pi v^2}{5},
\end{equation}
where $m_h$, $m_H$, and $m_5$ are the masses of $h$, $H$, and all the $H_5$ states, respectively.  Combining this with the sum rule in Eq.~(\ref{eq:kvhrule}) yields an upper bound on $\kappa_V^h$ as a function of the $H_5$ mass $m_5$,
\begin{equation}
	(\kappa_V^h)^2 < 1 + \frac{16 \pi v^2 - 5 m_h^2}{4 m_5^2 + 5 m_h^2}.
	\label{eq:highmass}
\end{equation}
This constraint becomes important for high $H_5$ masses above 700~GeV as shown in the left panel of Fig.~(\ref{fig1}).

These two constraints together set an absolute upper bound $\kappa_V^h \lesssim 1.57$ for $H_5$ mass greater than 100~GeV, which applies in all the generalized Georgi-Machacek models.

\begin{figure}
\resizebox{0.5\textwidth}{!}{\includegraphics{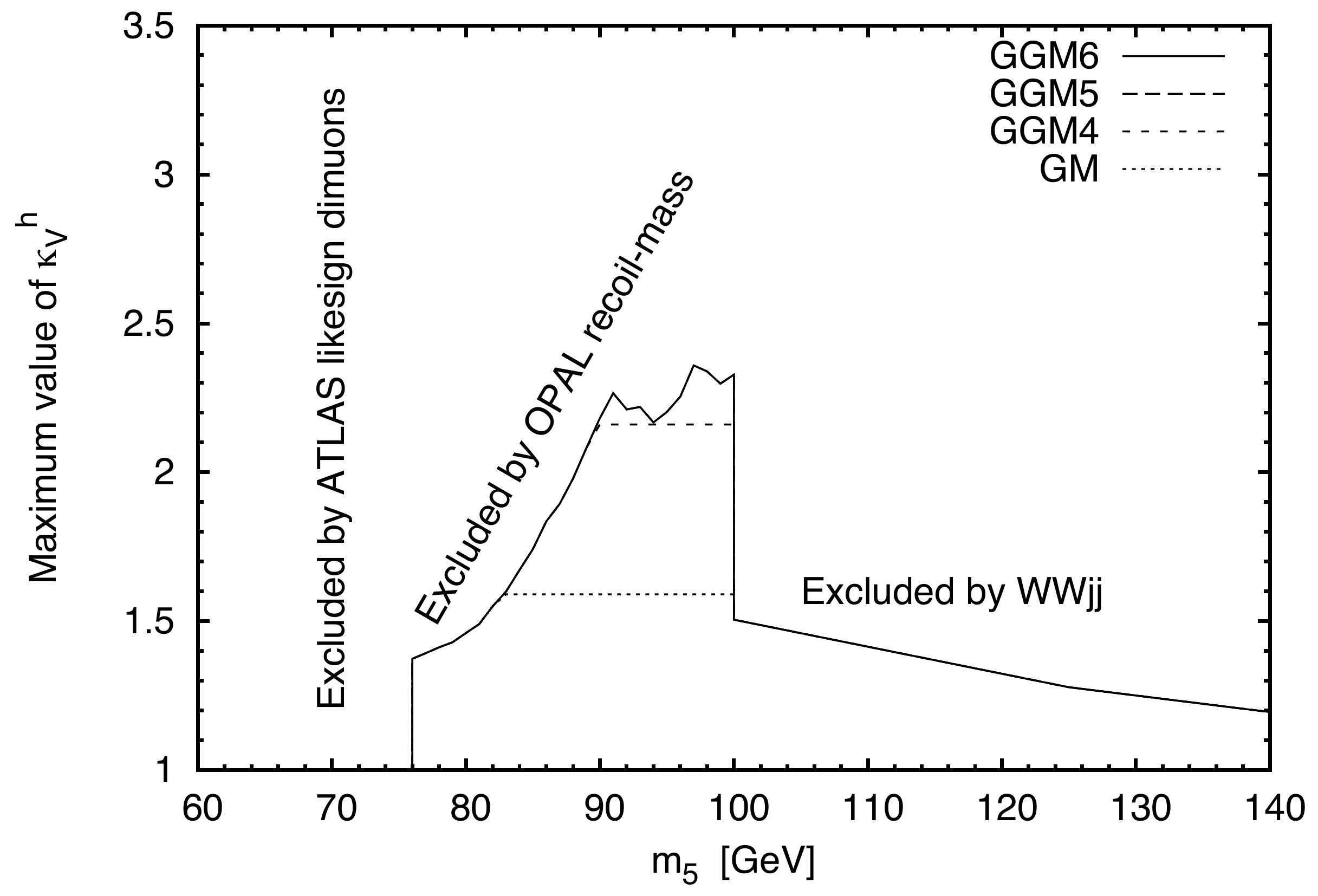}}%
\resizebox{0.5\textwidth}{!}{\includegraphics{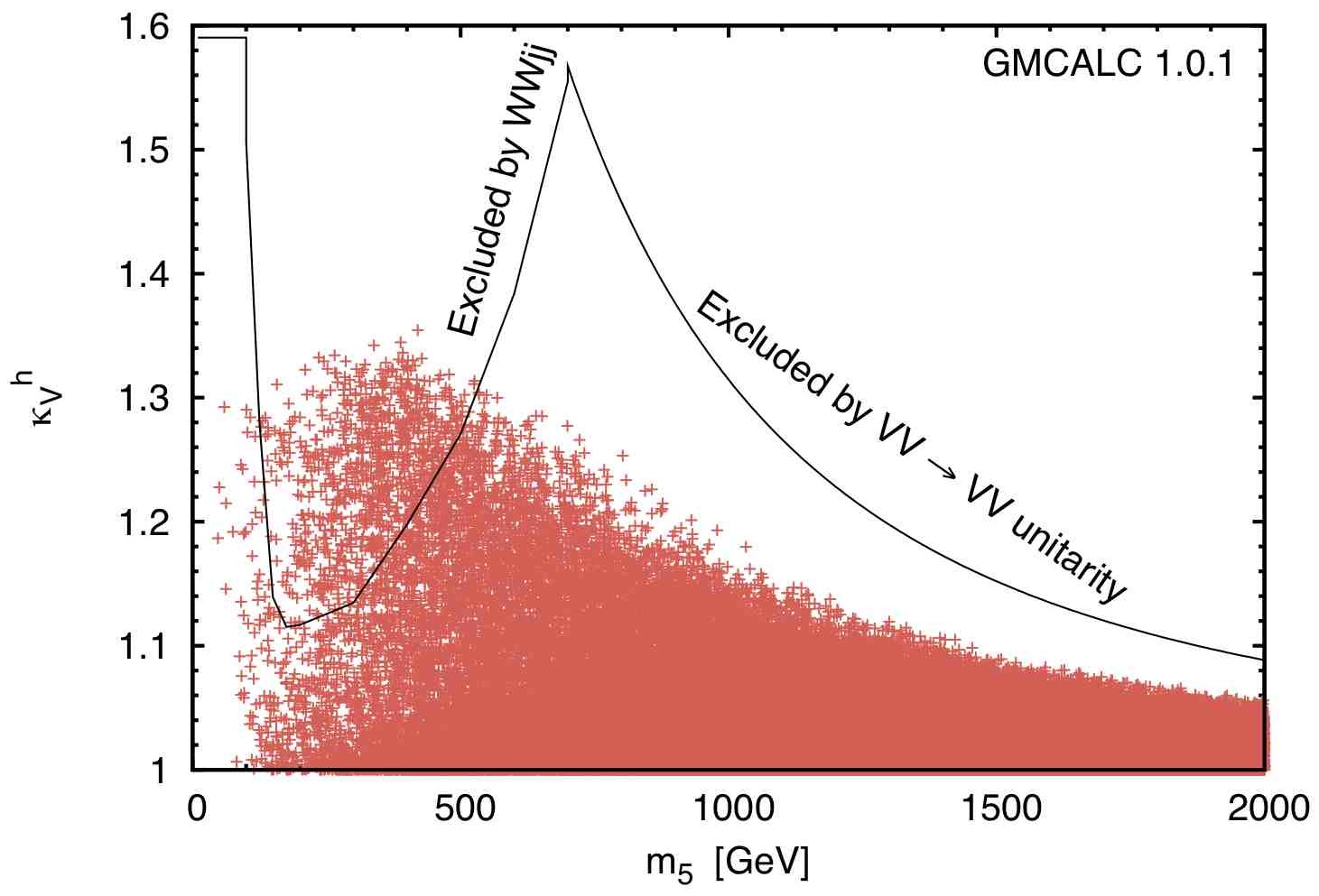}}
\caption{Left: Upper bound on $\kappa_V^h$ as obtained from the sum rule in Eq.~(\ref{eq:kvhrule}) as a function of the common $H_5$ mass $m_5$, focusing on the low mass region.  Constraints shown are the lower bound $m_5 > 76$~GeV from a recasting of Ref.~\cite{Kanemura:2014ipa} and a decay-mode-independent OPAL search for $ZH_5^0$~\cite{Abbiendi:2002qp}.  From Ref.~\cite{Logan:2015xpa}.
Right: As in Fig.~\ref{fig1} (left) for the original Georgi-Machacek model, but showing also the allowed points from a scan using GMCALC v1.0.1~\cite{Hartling:2014xma} after imposing all theoretical constraints and the constraint on $H_3$ from $b \to s\gamma$.
\label{fig2}}
\end{figure}

We now focus on the constraints for $m_5 < 100$~GeV.  Our third constraint comes from a recasting of an ATLAS search for prompt same-sign dimuon pairs~\cite{ATLAS:2014kca} into the context of the Higgs Triplet Model (HTM) assuming that the doubly-charged and singly-charged triplet states have the same mass~\cite{Kanemura:2014ipa}.  We in turn recast this result into the generalized Georgi-Machacek models (GM) by rescaling the $pp \to H^{++}H^{--}$ and $pp \to H^{\pm\pm}H^{\mp}$ cross sections by the appropriate $H_5 H_5 V$ gauge couplings, which are fixed by custodial symmetry to be the same in all the generalized Georgi-Machacek models.
This yields the dimuon cross sections after cuts shown in the right panel of Fig.~\ref{fig1}.  Applying the ATLAS measurement~\cite{ATLAS:2014kca} results in an absolute lower bound on the $H_5$ masses of 76~GeV, with no dependence on additional model parameters.  This lower bound is shown in the right panel of Fig.~\ref{fig2}.

Finally we obtain a fourth constraint from an OPAL search for $e^+e^- \to Z S^0$ production at the CERN Large Electron-Positron Collider (LEP), independent of the decay mode of the CP-even scalar $S^0$ using the recoil-mass method~\cite{Abbiendi:2002qp}.  The resulting limit has been conveniently implemented in HiggsBounds~\cite{HiggsBounds}.  This puts an upper bound on the $H_5^0ZZ$ coupling, and hence on $g_5$, as a function of the common $H_5$ mass; translating this into an upper bound on $\kappa_V^h$ results in the exclusion shown in the right panel of Fig.~\ref{fig2} for $m_5$ between 76 and 100~GeV.
Where no other constraints apply we have imposed the constraint $\tan\theta_H < 10/3$ from perturbativity of the top quark Yukawa coupling~\cite{Barger:1989fj}.

This leaves us with an absolute upper bound $\kappa_V^h \lesssim 2.36$ for any mass of $H_5$, which applies in all the generalized Georgi-Machacek models.

%%%%%%%%%%%%%%%%%%%%%%%%%%%%%%%%%%
\section{Outlook \label{sec:outlook}}

Custodial symmetry combined with unitarity sum rules provides a powerful tool to constrain Higgs sector extensions in which the $hVV$ coupling can be larger than in the SM by taking advantage of searches for the custodial fiveplet states $H_5^{\pm\pm}$, $H_5^{\pm}$, and $H_5^0$.  The weakest constraints are currently obtained for $H_5$ masses between 76 and 100~GeV; even so, an \emph{absolute} upper bound $\kappa_V^h \lesssim 2.36$ can be obtained, which holds in all the generalizations of the Georgi-Machacek model.

For high $H_5$ masses, the strongest constraint comes from tree-level perturbative unitarity of $VV \to VV$ scattering amplitudes.  Interestingly, the constraint given in Eq.~(\ref{eq:highmass}) is not fully saturated by the original Georgi-Machacek model, after imposing the full set of theoretical constraints.  This is shown in the right panel of Fig.~\ref{fig2}, where the red dots represent a scan using GMCALC v1.0.1~\cite{Hartling:2014xma}.  This suggests that a full analysis of the theoretical constraints on the scalar potentials for the generalized Georgi-Machacek models may further constrain the maximum enhancement of $\kappa_V^h$ in the high $H_5$ mass region.  We have written down these scalar potentials for the first time in Ref.~\cite{Logan:2015xpa}, but an analysis of their theoretical constraints is yet to be done.

We have not addressed here the septet extension of the SM, which also allows $\kappa_V^h > 1$.  The septet model does not preserve custodial symmetry in the scalar sector, so that there is no analogue of $H_5^0$ and the unitarity sum rules take a different form.  Constraints on $\kappa_V^h$ from searches for the doubly-charged scalar in the septet model are under investigation~\cite{Harris}.

%%%%%%%%%%%%%%%%%%%%%%%%%%%%%%%%%%
\begin{acknowledgments}
This talk was based on work done in collaboration with Vikram Rentala, Katy Hartling, Kunal Kumar, and Terry Pilkington, and supported by the Natural Sciences and Engineering Research Council of Canada.  I thank the organizers for their kind invitation to what has been a supremely enjoyable workshop and for partial financial support.
\end{acknowledgments}

\bigskip % extra skip inserted

\begin{thebibliography}{99} % Use for 10-99 references

\bibitem{Logan:2015xpa} 
  H.~E.~Logan and V.~Rentala,
  %``All the generalized Georgi-Machacek models,''
  arXiv:1502.01275 [hep-ph].
  %%CITATION = ARXIV:1502.01275;%%

\bibitem{Zeppenfeld:2000td} 
  D.~Zeppenfeld, R.~Kinnunen, A.~Nikitenko and E.~Richter-Was,
  %``Measuring Higgs boson couplings at the CERN LHC,''
  Phys.\ Rev.\ D {\bf 62}, 013009 (2000)
  [hep-ph/0002036];
  %%CITATION = HEP-PH/0002036;%%
%\bibitem{Djouadi:2000gu} 
  A.~Djouadi, R.~Kinnunen, E.~Richter-Was, H.~U.~Martyn, K.~A.~Assamagan, C.~Balazs, G.~Belanger and E.~Boos {\it et al.},
  %``The Higgs working group: Summary report,''
  hep-ph/0002258.
  %%CITATION = HEP-PH/0002258;%%
  
\bibitem{Kauer:2012hd} 
  N.~Kauer and G.~Passarino,
  %``Inadequacy of zero-width approximation for a light Higgs boson signal,''
  JHEP {\bf 1208}, 116 (2012)
  [arXiv:1206.4803 [hep-ph]];
  %%CITATION = ARXIV:1206.4803;%%
%\bibitem{Caola:2013yja} 
  F.~Caola and K.~Melnikov,
  %``Constraining the Higgs boson width with ZZ production at the LHC,''
  Phys.\ Rev.\ D {\bf 88}, 054024 (2013)
  [arXiv:1307.4935 [hep-ph]];
  %%CITATION = ARXIV:1307.4935;%%  
%\bibitem{Campbell:2013una} 
  J.~M.~Campbell, R.~K.~Ellis and C.~Williams,
  %``Bounding the Higgs width at the LHC using full analytic results for $gg -> e^- e^+ \mu^- \mu^+$,''
  JHEP {\bf 1404}, 060 (2014)
  [arXiv:1311.3589 [hep-ph]].
  %%CITATION = ARXIV:1311.3589;%%
  
\bibitem{Logan:2014ppa} 
  H.~E.~Logan,
  %``Hiding a Higgs width enhancement from off-shell gg (--> h*) --> ZZ measurements,''
  arXiv:1412.7577 [hep-ph].
  %%CITATION = ARXIV:1412.7577;%%

\bibitem{Lee:1977yc} 
  B.~W.~Lee, C.~Quigg and H.~B.~Thacker,
  %``The Strength of Weak Interactions at Very High-Energies and the Higgs Boson Mass,''
  Phys.\ Rev.\ Lett.\  {\bf 38}, 883 (1977);
  %%CITATION = PRLTA,38,883;%%
%\bibitem{Lee:1977eg} 
%  B.~W.~Lee, C.~Quigg and H.~B.~Thacker,
  %``Weak Interactions at Very High-Energies: The Role of the Higgs Boson Mass,''
  Phys.\ Rev.\ D {\bf 16}, 1519 (1977).
  %%CITATION = PHRVA,D16,1519;%%

\bibitem{Gunion:1990kf} 
  J.~F.~Gunion, H.~E.~Haber and J.~Wudka,
  %``Sum rules for Higgs bosons,''
  Phys.\ Rev.\ D {\bf 43}, 904 (1991).
  %%CITATION = PHRVA,D43,904;%%

\bibitem{Falkowski:2012vh} 
  A.~Falkowski, S.~Rychkov and A.~Urbano,
  %``What if the Higgs couplings to W and Z bosons are larger than in the Standard Model?,''
  JHEP {\bf 1204}, 073 (2012)
  [arXiv:1202.1532 [hep-ph]];
  %%CITATION = ARXIV:1202.1532;%%
  %\bibitem{Grinstein:2013fia} 
  B.~Grinstein, C.~W.~Murphy, D.~Pirtskhalava and P.~Uttayarat,
  %``Theoretical Constraints on Additional Higgs Bosons in Light of the 126 GeV Higgs,''
  JHEP {\bf 1405}, 083 (2014)
  [arXiv:1401.0070 [hep-ph]];
  %%CITATION = ARXIV:1401.0070;%%
%\bibitem{Bellazzini:2014waa} 
  B.~Bellazzini, L.~Martucci and R.~Torre,
  %``Symmetries, Sum Rules and Constraints on Effective Field Theories,''
  JHEP {\bf 1409}, 100 (2014)
  [arXiv:1405.2960 [hep-th]].
  %%CITATION = ARXIV:1405.2960;%%

\bibitem{Gunion:1989we} 
  J.~F.~Gunion, H.~E.~Haber, G.~L.~Kane and S.~Dawson,
  {\it The Higgs Hunter's Guide}
(Westview, Boulder, Colorado, 2000).  
%  Front.\ Phys.\  {\bf 80}, 1 (2000).
  %%CITATION = FRPHA,80,1;%%

\bibitem{Hally:2012pu} 
  K.~Hally, H.~E.~Logan and T.~Pilkington,
  %``Constraints on large scalar multiplets from perturbative unitarity,''
  Phys.\ Rev.\ D {\bf 85}, 095017 (2012)
  [arXiv:1202.5073 [hep-ph]].
  %%CITATION = ARXIV:1202.5073;%%

\bibitem{Agashe:2014kda} 
  K.~A.~Olive {\it et al.}  [Particle Data Group Collaboration],
  %``Review of Particle Physics,''
  Chin.\ Phys.\ C {\bf 38}, 090001 (2014).
  %%CITATION = CHPHD,C38,090001;%%  
  
\bibitem{Hisano:2013sn} 
  J.~Hisano and K.~Tsumura,
  %``Higgs boson mixes with an SU(2) septet representation,''
  Phys.\ Rev.\ D {\bf 87}, 053004 (2013)
  [arXiv:1301.6455 [hep-ph]].
  %%CITATION = ARXIV:1301.6455;%%
  
\bibitem{Kanemura:2013mc} 
  S.~Kanemura, M.~Kikuchi and K.~Yagyu,
  %``Probing exotic Higgs sectors from the precise measurement of Higgs boson couplings,''
  Phys.\ Rev.\ D {\bf 88}, 015020 (2013)
  [arXiv:1301.7303 [hep-ph]].
  %%CITATION = ARXIV:1301.7303;%%
  
\bibitem{Georgi:1985nv} 
  H.~Georgi and M.~Machacek,
  %``Doubly Charged Higgs Bosons,''
  Nucl.\ Phys.\ B {\bf 262}, 463 (1985);
  %%CITATION = NUPHA,B262,463;%%
%\bibitem{Chanowitz:1985ug} 
  M.~S.~Chanowitz and M.~Golden,
  %``Higgs Boson Triplets With M ($W$) = M ($Z$) $\cos \theta \omega$,''
  Phys.\ Lett.\ B {\bf 165}, 105 (1985).
  %%CITATION = PHLTA,B165,105;%%
 
\bibitem{Gunion:1990dt} 
  J.~F.~Gunion, R.~Vega and J.~Wudka,
  %``Naturalness problems for rho = 1 and other large one loop effects for a standard model Higgs sector containing triplet fields,''
  Phys.\ Rev.\ D {\bf 43}, 2322 (1991).
  %%CITATION = PHRVA,D43,2322;%% 

\bibitem{Garcia-Pepin:2014yfa} 
  M.~Garcia-Pepin, S.~Gori, M.~Quiros, R.~Vega, R.~Vega-Morales and T.~T.~Yu,
  %``Supersymmetric Custodial Higgs Triplets and the Breaking of Universality,''
  Phys.\ Rev.\ D {\bf 91}, 015016 (2015)
  [arXiv:1409.5737 [hep-ph]].
  %%CITATION = ARXIV:1409.5737;%%  

\bibitem{Hartling:2014zca} 
  K.~Hartling, K.~Kumar and H.~E.~Logan,
  %``The decoupling limit in the Georgi-Machacek model,''
  Phys.\ Rev.\ D {\bf 90}, 015007 (2014)
  [arXiv:1404.2640 [hep-ph]].
  %%CITATION = ARXIV:1404.2640;%%

\bibitem{Galison:1983qg} 
  P.~Galison,
  %``Large Weak Isospin and the $W$ Mass,''
  Nucl.\ Phys.\ B {\bf 232}, 26 (1984);
  %%CITATION = NUPHA,B232,26;%%
 %\bibitem{Robinett:1985ec} 
  R.~W.~Robinett,
  %``Extended Strongly Interacting Higgs Theories,''
  Phys.\ Rev.\ D {\bf 32}, 1780 (1985);
  %%CITATION = PHRVA,D32,1780;%%
%\bibitem{Logan:1999if} 
  H.~E.~Logan,
  %``Radiative corrections to the Z b anti-b vertex and constraints on extended Higgs sectors,''
  hep-ph/9906332;
  %%CITATION = HEP-PH/9906332;%%
 %\bibitem{Chang:2012gn} 
  S.~Chang, C.~A.~Newby, N.~Raj and C.~Wanotayaroj,
  %``Revisiting Theories with Enhanced Higgs Couplings to Weak Gauge Bosons,''
  Phys.\ Rev.\ D {\bf 86}, 095015 (2012)
  [arXiv:1207.0493 [hep-ph]].
  %%CITATION = ARXIV:1207.0493;%%

\bibitem{Barger:1989fj} 
  V.~D.~Barger, J.~L.~Hewett and R.~J.~N.~Phillips,
  %``New Constraints on the Charged Higgs Sector in Two Higgs Doublet Models,''
  Phys.\ Rev.\ D {\bf 41}, 3421 (1990).
  %%CITATION = PHRVA,D41,3421;%%
  
 \bibitem{Hartling:2014aga} 
  K.~Hartling, K.~Kumar and H.~E.~Logan,
  %``Indirect constraints on the Georgi-Machacek model and implications for Higgs boson couplings,''
  Phys.\ Rev.\ D {\bf 91}, 015013 (2015)
  [arXiv:1410.5538 [hep-ph]].
  %%CITATION = ARXIV:1410.5538;%%
  
\bibitem{ATLASWWjj} 
  The ATLAS collaboration,
  %``Evidence for electroweak production of $W^{\pm}W^{\pm}jj$ in $pp$ collisions at $\sqrt{s}=8$ TeV with the ATLAS detector,''
  ATLAS-CONF-2014-013, ATLAS-COM-CONF-2014-015.
  %%CITATION = ATLAS-CONF-2014-013, ATLAS-COM-CONF-2014-015;%%
  
\bibitem{Chiang:2014bia} 
  C.~W.~Chiang, S.~Kanemura and K.~Yagyu,
  %``Novel constraint on the parameter space of the Georgi-Machacek model with current LHC data,''
  Phys.\ Rev.\ D {\bf 90}, 115025 (2014)
  [arXiv:1407.5053 [hep-ph]].
  %%CITATION = ARXIV:1407.5053;%%  
  
\bibitem{ATLAS:2014kca} 
  G.~Aad {\it et al.}  [ATLAS Collaboration],
  %``Search for anomalous production of prompt same-sign lepton pairs and pair-produced doubly charged Higgs bosons with $ \sqrt{s}=8 $ TeV $pp$ collisions using the ATLAS detector,''
  JHEP {\bf 1503}, 041 (2015)
  [arXiv:1412.0237 [hep-ex]].
  %%CITATION = ARXIV:1412.0237;%%
  
  \bibitem{Kanemura:2014ipa} 
  S.~Kanemura, M.~Kikuchi, H.~Yokoya and K.~Yagyu,
  %``LHC Run-I constraint on the mass of doubly charged Higgs bosons in the same-sign diboson decay scenario,''
  arXiv:1412.7603 [hep-ph].
  %%CITATION = ARXIV:1412.7603;%%

\bibitem{Abbiendi:2002qp} 
  G.~Abbiendi {\it et al.}  [OPAL Collaboration],
  %``Decay mode independent searches for new scalar bosons with the OPAL detector at LEP,''
  Eur.\ Phys.\ J.\ C {\bf 27}, 311 (2003)
  [hep-ex/0206022].
  %%CITATION = HEP-EX/0206022;%%
  
\bibitem{HiggsBounds}
%\bibitem{Bechtle:2008jh}
  P.~Bechtle, O.~Brein, S.~Heinemeyer, G.~Weiglein and K.~E.~Williams,
  %``HiggsBounds: Confronting Arbitrary Higgs Sectors with Exclusion Bounds from LEP and the Tevatron,''
  Comput.\ Phys.\ Commun.\  {\bf 181}, 138 (2010)
  [arXiv:0811.4169 [hep-ph]];
  %%CITATION = ARXIV:0811.4169;%%
%\bibitem{Bechtle:2011sb}
 % P.~Bechtle, O.~Brein, S.~Heinemeyer, G.~Weiglein and K.~E.~Williams,
  %``HiggsBounds 2.0.0: Confronting Neutral and Charged Higgs Sector Predictions with Exclusion Bounds from LEP and the Tevatron,''
  Comput.\ Phys.\ Commun.\  {\bf 182}, 2605 (2011)
  [arXiv:1102.1898 [hep-ph]];
  %%CITATION = ARXIV:1102.1898;%%
%\bibitem{Bechtle:2013gu}
  P.~Bechtle, O.~Brein, S.~Heinemeyer, O.~Stal, T.~Stefaniak, G.~Weiglein and K.~E.~Williams,
  %``Recent Developments in HiggsBounds and a Preview of HiggsSignals,''
  PoS CHARGED {\bf 2012}, 024 (2012)
  [arXiv:1301.2345 [hep-ph]];
  %%CITATION = ARXIV:1301.2345;%%
%\bibitem{Bechtle:2013wla}
%  P.~Bechtle, O.~Brein, S.~Heinemeyer, O.~StŒl, T.~Stefaniak, G.~Weiglein and K.~E.~Williams,
  %``$\mathsf{HiggsBounds}-4$: Improved Tests of Extended Higgs Sectors against Exclusion Bounds from LEP, the Tevatron and the LHC,''
  Eur.\ Phys.\ J.\ C {\bf 74}, 2693 (2014)
  [arXiv:1311.0055 [hep-ph]].
  %%CITATION = ARXIV:1311.0055;%%

\bibitem{Hartling:2014xma} 
  K.~Hartling, K.~Kumar and H.~E.~Logan,
  %``GMCALC: a calculator for the Georgi-Machacek model,''
  arXiv:1412.7387 [hep-ph].
  %%CITATION = ARXIV:1412.7387;%%
  
\bibitem{Harris}
M.-J.~Harris and H.~E.~Logan, in preparation.
   
\end{thebibliography}

\end{document}